\documentclass[12pt, preprint]{aastex}

\begin{document}

\title{Subaru Spectroscopy of the Giant Ly$\alpha$ Nebula Associated with 
the High-$z$ Powerful Radio Galaxy 1243+036\footnote{Based on
data collected at Subaru Telescope, which is operated by the National
Astronomical Observatory of Japan.}
}

\author{Youichi Ohyama$^2$, Yoshiaki Taniguchi$^3$}

\affil{$^2$Subaru Telescope, National Astronomical Observatory
              of Japan, 650 N. A`ohoku Place, University Park, Hilo,
              HI 96720 \\
     $^3$Astronomical Institute, Graduate School of Science,
           Tohoku University, Aramaki, Aoba, Sendai 980-8578, Japan}

\begin{abstract}

We report results of our new spatially-resolved, optical spectroscopy of the 
giant Ly$\alpha$ nebula around a powerful radio galaxy 1243+036
(4C+03.24) at $z=3.57$.
The nebula is extended over $\sim 30$ kpc from the nucleus, and forms a pair
of cones or elongated bubbles.
The high-velocity ($\sim -1000$ km s$^{-1}$; blueshifted with respect to the
systemic velocity) Ly$\alpha$-emitting components are detected at both sides
of the nucleus along its major axis.
The northwestern nebula is more spectacular in its velocity shift (blueshifted
by $-1000$ km s$^{-1}$ to $-1400$ km s$^{-1}$) and in its width ($\simeq 1900$
km s$^{-1}$ FWHM) over $\simeq 30$ kpc scale.
We discuss possible origin of the nebula; 1) the shock-heated
expanding bubble or outflowing cone associated with the superwind activity of
the host galaxy,
2) halo gas photoionized by the anisotropic radiation from the active galactic
nuclei (AGN), and 3) the jet-induced star-formation or shock.
The last possibility may not be likely because Ly$\alpha$ emission is distributed
out of the narrow channel of the radio jet.
We show that the superwind model is most plausible since it can
explain both the characteristics of the morphology (size and shape)
and the kinematical structures (velocity shift and line width) of the
nebula although the photoionization by AGN may contribute to the
excitation to some extent.

\end{abstract}


\section{INTRODUCTION}

It is well known that images of the rest-frame UV and optical continua
are elongated preferentially along the radio axis in powerful radio
galaxies (PRGs) at redshift ($z$) $> 0.6$ (e.g., Chambers et al. 1987;
McCarthy et al. 1987); the so-called alignment effect.
Indeed, many high-$z$ ($z>2$) PRGs (HzPRGs) show the 
alignment effect, and its origin has been in debate in this decade
(e.g., McCarthy 1993).
Various models have been proposed to explain the
alignment effect; e.g., (1) scattering of the anisotropic
radiation from a central engine of active galactic nuclei (AGN: e.g.,
di Serego Alighieri et al. 1989; Fabian 1989; Cimatti et al. 1998),
(2) the jet-induced star formation (e.g.,
Chambers et al. 1987; McCarthy et al. 1987), 
and (3) the jet-induced shock heating (e.g., De Breuck et al. 2000).
In these models, AGNs play some
important roles to create an extended aligned continuum
emission at various wavelengths.
The nebula contribution to the extended UV continua is also suggested
in some PRGs (e.g., Dickson et al. 1995).

An alternative idea, a pair of elongated bubbles or bi-directional
cones associated with the superwind/starburst activity of the host
galaxy, is proposed to explain a figure-8 shaped extended
nebula in another radio galaxy, MRC 0406-244, at $z=2.4$ (Taniguchi et al.
2001; see also Rush et al. 1997; Pentericci et al. 2001).
Although the radio-jet activity and
the starburst/superwind activity are different physical processes
and their timescales are also significantly different, the alignment effect 
can be understood in the following way (Taniguchi et al. 2001).
The superwind could blow as a bipolar wind, which is often observed in many
superwind galaxies in the local universe (see Heckman et al. 1990),
since it is likely that the gas
in the host galaxy is distributed with a disk-like configuration even for
a young host galaxy of HzPRGs.
The radio jet is also expelled to the two directions
perpendicular to the accretion plasma disk.
Therefore, we can explain the alignment effect if
the accretion disk is nearly co-planer to the host gas disk.
Since some of these objects are known to experience vigorous star formation
activity (e.g., Dey et al. 1997), this new mechanism can be regarded as
one of possible ideas.

Study of such superwind nebulae in high-$z$ universe seems very useful 
to understand the formation and evolution of galaxies from a general
point of view. The reason for this is that
galaxies are expected to experience the galactic wind (i.e.,
the powerful initial superwind in a late phase of the galaxy formation),
which is considered to play significant roles
to determine global characteristics of the present-day massive elliptical
galaxies (e.g., Arimoto \& Yoshii 1987).
Since we know many HzPRGs even at redshift $z \gtrsim$ 3 (e.g.,
R\"ottgering 1997; van Ojik et al. 1997; van Breugel et al. 1999,
De Breuck et al. 2002)
thanks to their powerful radio emission, their superwind activities can be
used to probe early star formation history in their host galaxies.
Powerful 8-10 m-class telescopes enable us to investigate them in detail
(e.g., Villar-Martin et al. 2000).

In this paper, we focus our attention to one of HzPRGs, 1243+036
(= 4C+03.24), at $z=3.6$, which has the aligned Ly$\alpha$ nebula,
and is one of the highest-$z$ objects known to show an alignment effect
(see, e.g., Chambers et al. 1990; Lacy et al. 1994).
This object was identified as a HzPRG during the course of spectroscopic 
survey of ultra steep spectrum sources conducted at ESO
(R\"ottgering 1994, 1995, 1997; van Ojik et al. 1997).
Its detailed follow-up observations were made by van Ojik et al. 
(1996; hereafter vO96).
According to vO96, the observational properties of
the nebula are very complex in morphology and kinematics,
and can be summarized as follows\footnote{
In this paper, we assume a Hubble constant of $H_{\rm 0}=50$ km
s$^{-1}$ Mpc$^{-1}$ and a deceleration parameter of $q_{\rm 0}=0.5$
for consistency with the previous study of vO96.
With these cosmological parameters, 1 arcsecond corresponds to 6.8 kpc
at a distance of 1243+036.}. (a) This object is surrounded by
a huge Ly$\alpha$ halo which extends over 135 kpc.
(b) Its Ly$\alpha$ luminosity amounts to $L$(Ly$\alpha$) $\simeq 10^{44.5}
h_{0.5}^{-2}$ ergs s$^{-1}$. (c) Its morphology shows the alignment
effect. (d) Its kinematical properties show the presence of the following
two components; i) the high-velocity width component with
FWHM (full width at half maximum) $\simeq$ 1500 km s$^{-1}$ at
$-1100$ km s$^{-1}$ (blueshifted), and ii)
the halo component with FWHM $\simeq$ 250 km s$^{-1}$.
The halo component extends over 135 kpc
with an average velocity gradient of $\simeq$ 450 km s$^{-1}$/135 kpc.
In addition, recent high-resolution images of $R$-band, $K_{\rm s}$-band,
and narrow-band near 2.3$\mu$m (corresponding to the rest-frame UV-optical
continua and redshifted [OIII]4959, 5007\AA\AA~ emission line, respectively)
have revealed closely-aligned continuum and line emission along the narrow
channel of the radio jet (van Breugel et al. 1998; Pentericci et al. 1999).
Although the close spatial coincidence between the UV/optical and radio
emission strongly suggests a possibility of the radio jet-induced alignment
component in this object, the fan- or cone-like appearance of the Ly$\alpha$
nebula seems to indicate the presence of another nebula component
within the same object.
Therefore it is interesting to investigate 
this object in more detail and to explore its origin unambiguously.

\section{OBSERVATION AND DATA REDUCTION}

Observation was made with the Faint Object Camera And Spectrograph,
FOCAS (Kashikawa et al. 2002), attached at
the Cassegrain focus of the Subaru telescope (Kaifu 1998) on 
April 22, 2002 (UT) to study the Ly$\alpha$ nebula around 1243+036.
The sky condition was clear and the seeing size (FWHM) was 0.\arcsec6.
A 300 second $V$-band image, which includes the redshifted Ly$\alpha$
emission at $\simeq 5560$\AA, was obtained without CCD binning (0.\arcsec1~
per pixel) just before the spectroscopic observation for target acquisition
purpose.
Then, a 0.\arcsec8~ long slit was placed on the peak of the $V$-band
image at a position angle (PA) of 152$^\circ$.
The slit positioning accuracy was estimated to be about 0.\arcsec5~
due to the faintness of the target.
$3\times 2$ pixel$^2$ CCD binning was applied at the time of
spectroscopy, and a single 1800 sec exposure was taken to give a
spectrum covering from 4700\AA~ to 9400\AA~ with the combination of
both the medium-dispersion grism (300B) with 300 grooves mm$^{-1}$ and
the order sorting filter (Y47).
The resultant resolution was 0.\arcsec3~ in space and 2.81\AA~ in
wavelength per pixel corresponding 150 km s$^{-1}$ at the redshifted
Ly$\alpha$ emission.
The data reduction was made by using both the IDL-based FOCAS data
reduction software (Yoshida et al. 2000) and the IRAF in a standard
way (overscan subtraction, bias subtraction, and flat fielding by
using the dome flat images).
The wavelength calibration was made by using sky emission lines, and
could be done with high accuracy ($\simeq 6$\AA)
because the redshifted Ly$\alpha$
emission is very close to the strong [O {\sc i}] sky emission at 5577\AA.

\section{RESULTS}

\subsection{Overall Morphology of the Ly$\alpha$ Nebula}

We show our $V$-band image overlaid on the continuum image
taken with a F702W filter of HST (Pentericci et al. 1999) in Figure 1.
The $V$-band image is smoothed by Gaussian convolution to enhance the weak
emission at outer parts of the nebula; the effective seeing (after
the smoothing) is 1.\arcsec1~ FWHM.
The total magnitude of the nebula is $V=22.6\pm 0.3$.
It is clearly shown that our $V$-band image reveals widely extended
nebula component, showing remarkable contrast with the continuum emission,
which is tightly aligned to the narrow radio jet (vO96; Pentericci et al. 1999).
Note also that the $V$-band image appears very similar to the Ly$\alpha$
image taken with a narrow-band filter (vO96)
and thus our $V$-band imaging traces the Ly$\alpha$ mostly.
However, there is a notable difference between the two at northwest (NW) of
the nucleus:
Although the Ly$\alpha$ image shows an extended Ly$\alpha$ emission only toward
southeast (SE) of the nucleus, our $V$-band image shows an extended nebula toward
both NW and SE of the nucleus.
Since our $V$-band filter bandpass covers the whole redshifted Ly$\alpha$ emission,
and the contribution of the continuum emission should be almost negligible for
the extended component judging from both the prominent nuclear source and
weak emission extended along the jet (vO96; van Breugel et al. 1998;
Pentericci et al. 1999), our $V$-band image and the Ly$\alpha$ image of vO96
should reveal the gaseous nebula in a similar way.
Therefore this apparent discrepancy in two images needs an explanation.

Our Ly$\alpha$ spectrogram is shown in Figure 2 along with the slit position
overlaid on our $V$-band image.
The spectrogram is also smoothed by 4 by 4 binned-pixel running averaging to
enhance the outer weak emission.
After the smoothing, the effective resolution is 1.\arcsec2~ in space, which is
almost comparable to that of the smoothed imaging data.
The velocity resolution is 600 km s$^{-1}$.
In order to check the consistency quantitatively between the spectrogram and the
$V$-band image,
we made an one-dimensional plot of the spatial Ly$\alpha$ flux distribution by
integrating the $V$-band flux over an expected slit position along the slit width
directions (Figure 2b).
A similar plot was also made from the spectroscopic data by
integrating the flux on the spectrogram along the wavelength direction
at each spatial position (Figure 2d).
These two plots look very similar to each other; i.e.,
both show not only an asymmetric distribution toward SE but also a
diffuse extension toward NW at a fainter flux level.
This suggests not only that our spectroscopy and imaging data are consistent with
each other (i.e., the spatial orientation of the spectrogram is right, and the
slit was placed at the expected position at the center of the nebula),
but also that the NW nebula does exist.
It seems noteworthy that the NW component is blueshifted at 5541\AA~ or bluer
for the most part (Figure 2c).
Such a blueshifted component could not be detected by vO96 due to rather narrow
wavelength coverage of their narrow-band filter (5541\AA~ -- 5601\AA).
Therefore, we conclude that there is a pair of elongated nebulae along NW and SE
directions around 1243+036.

\subsection{Description of Ly$\alpha$ Spectrogram}

Our Ly$\alpha$ spectrogram (Figure 2c)
reveals complicated structures of the nebula both in space and in
velocity.
We group the nebula into the following four components:
(I) the compact nucleus component, (II) the extended near-systemic
component at both sides of the nucleus, (III) the redshifted component
aside the nucleus, and (IV) the extended blue components 
at both sides of the nucleus.
We give below a summary of their observational properties and their
comparisons with results of vO96, who conducted similar long-slit spectroscopy.
Note that, although the slit position (centered on the nebula) and its
orientation (PA=152$^\circ$) are the same for the two, the
slit widths (2.\arcsec5~ for vO96 and 0.\arcsec8~ for us) and the
velocity resolutions (150 km s$^{-1}$ for vO96 and $600$ km s$^{-1}$
for us) are different.

(I) The component at $< \pm 1$\arcsec~ around the peak of the Ly$\alpha$
emission shows an evident blue-asymmetric profile with its peak at
5560\AA~, being consistent with the report by vO96 (at 5557\AA~).
Its intensity peak corresponds spatially to that of 
the rest UV-optical continuum emission with which the radio core
showing flatter spectrum is associated (vO96; van Breugel et
al. 1998; Pentericci et al. 1999). Therefore, we attribute this component to the
nucleus component.
The blue wing emission can be traced down to $-1400$ km s$^{-1}$ in the rest
frame of the galaxy (hereafter all velocities are shown with respect
to the systemic velocity in the rest frame of the galaxy) although a
part of the wing may be contaminated by the off-nucleus blue
components at both sides of the nucleus (see below).

(II) The extended component whose velocity is close to the systemic one
 is found at both sides of the
nucleus, and can be traced out to $\pm$ (4-5) \arcsec~ from the nucleus.
It shows no strong velocity shear across the extension ($< 600$ km
s$^{-1}$) although the sky emission at 5577\AA~ might affect
the measurement of its velocity curve given our lower velocity
resolution.
The velocity width is narrower than the nucleus component across
the extension since we could not resolved the width by our instrument
resolution (600 km s$^{-1}$ FWHM).

vO96 detected an ``outer halo'' component extending out to $\pm 10$\arcsec~
on both sides of the nucleus at almost systemic velocity.
This component has a narrow line width ($\simeq 250$ km s$^{-1}$ FWHM)
and shows a large-scale velocity shear of as much as $\pm 225$ km s$^{-1}$.
Although our spectroscopic resolution is not high enough to investigate
such a velocity curve, 
our near-systemic component appears to correspond to
the outer halo component of vO96 although we detect it only out to
$\pm 4-5$\arcsec.

(III) The redshifted component is detected at $\simeq 1$\arcsec~ NW
of the nucleus
\footnote{
Note that similar redshifted component is detected at $\simeq 1$\arcsec~
SE of the nucleus (i.e., another side of the nucleus) in Figure 11 of vO96.
}, and is found at redward of the sky emission.
Due to the strong sky contamination, little information is available
for this component.

(IV) The blueshifted component at SE extends up to $2$\arcsec~ from
the nucleus and shows strong blue-shifted velocity of $\simeq -1000$ km
s$^{-1}$.
At the opposite side of the nucleus (NW), another blue component is
detected which is much more extended ($\simeq 4$\arcsec~ -- $5$\arcsec~ from
the nucleus) and shows much wider velocity width ($\simeq 1900$ km s$^{-1}$
FWHM corrected for the instrumental resolution\footnote{
Hereafter all line widths are given by
a FWHM corrected for the instrumental resolution.}).
The velocity curve shows the bluest velocity of $-1400$ km s$^{-1}$ at
$\simeq$ 2.\arcsec5~ NW, and the velocity slows down at outer regions
($\simeq -1000$ km s$^{-1}$ at $\simeq 5$\arcsec~ NW).

vO96 has also detected blueshifted and extended component.
Note, however, that there might be a difference between their spectrogram
and ours.
We could not detect any compact emission-line component at $\sim 2$\arcsec~
NW from the nucleus
\footnote{
We found that the spectrogram presented in vO96 (their Figure 11) is
probably shown in
wrong spatial direction (i.e., the spatial axis is flipped around the nucleus).
This idea is based on the facts that
(1) the redshifted component (III) is found at NW of the nucleus
in our spectrogram, rather than at SE of the nucleus at a similar velocity,
and
(2) our $V$-band image and spectrogram are consistent with each other
in orientation, as shown in section 3.1, and our $V$-band image and their
narrow-band Ly$\alpha$ image (their Figure 9) are also consistent
with each other in orientation, given the different wavelength coverage
of the nebula (as explained in section 3.1),
ensuring that our orientation is right.
Therefore, although vO96 claimed that this high-velocity component is
found at SE, its correct position is probably at the opposite side (NW) of
the nucleus.
In the following, we assume an image flip of the vO96's spectrogram for
the comparison.
},
although vO96 claimed a presence of an enhanced Ly$\alpha$ emission there
which shows blueshifted velocity of $-1100$ km s$^{-1}$
with a wider line width of $1200$ km s$^{-1}$ FWHM.
Our spectrogram shows, on the other hand, a component which shows similar
kinematical properties (blueshifted by $-1000 - -1400$ km s$^{-1}$
with an even wider line width of $1900$ km s$^{-1}$ FWHM)
but is extended more smoothly over $\simeq 4$\arcsec.
This difference could be explained by the different slit coverage on the
nebula between the two spectrograms, i.e., our narrow slit ($\simeq 1/3$ of
vO96's) could not cover a portion of the nebula where enhanced Ly$\alpha$ emission
is present.
If this is the case, the nebula might have complex sub-structures within a nebula.

\section{DISCUSSION}

\subsection{Origin of the Ly$\alpha$ nebula}

We discuss plausible origins of the alignment effect observed in 1243+036.
There are some pieces of evidence indicating that other mechanism
than the radio-jet related one works to excite the extended nebula in 1243+036.
First, the previous high-resolution images of $R$-band, $K_{\rm s}$-band,
and narrow-band near 2.3$\mu$m (corresponding to the rest-frame UV-optical
continua and redshifted [OIII]4959, 5007\AA\AA~ emission line, respectively)
have revealed closely-aligned continuum and line emission along the narrow
channel of the radio jet (van Breugel et al. 1998; Pentericci et al. 1999).
Although this indicates an importance of the radio jet activity
to the alignment effect, the Ly$\alpha$ nebula of this galaxy is extended
much more widely than the region of the aligned continuum (Figure 1).
Second, the extended components are blueshifted at both side of the nucleus.
The jet-induced extranuclear shock, if any, should show a pair of blueshifted
and redshifted components at each side of the nucleus because the symmetric
propagation of a pair of jet and its counter jet is expected,
being inconsistent with the observation.
Therefore it seems very likely that the effect of the radio jet onto
the extended nebula can not explain the excitation of the whole Ly$\alpha$
nebula.

Another idea of the origin of the widely extended Ly$\alpha$ nebula is
the circumgalactic matter ionized by the anisotropic hard radiation
from AGN.
If this is the case, the matter irradiated by the AGN radiation is required to
be distributed over a large scale ($\simeq 60$ kpc) at a violent
kinematical status to explain the velocity shift of $\simeq -1000$ km s$^{-1}$
together with a large velocity dispersion of 1000 - 2000 km s$^{-1}$.
One possibility may be that such violent gaseous systems are
tidally-induced structures formed through a putative merger event.
However, even nearby luminous mergers 
show an overall velocity differences of $\simeq 600$ km s$^{-1}$ or less
along the tidal tails (e.g., Hibbard \& Yun 1999; Mihos \& Bothun 1998).
Also the wider velocity dispersion seems difficult to be reproduced along
the tidal tail (e.g., Hibbard \& Yun 1999; Mihos \& Bothun 1998).
Another candidate for the extended matter is the remaining sub-galactic clumps
for the galaxy formation.
However, the range of the clump velocity dispersion would be 300 km
s$^{-1}$ or less
because gaseous clumps with a higher velocity dispersion are difficult
for them to assemble to form a galaxy.
Therefore this idea could not be simply applied for the case of 1243+036.

An alternative idea for the origin of the blue-shifted extended
Ly$\alpha$ nebula is the shock-excited nebula associating with the superwind
outflow.
Our $V$-band image reveals a pair of elongated bubbles or cone-like
structures extending into opposite ways from the nucleus.
Similar extended nebulae are often seen around the starburst galaxies
with the superwind/superbubble activity, such as M82, NGC 3079, and Arp 220
(e.g., Heckman et al. 1990).
Also similar nebula is found around a high-$z$ ($z=2.4$) radio galaxy
MRC 0406-244 which shows a pair of shock-excited elongated bubbles
associated with the superwind activity (Taniguchi et al. 2001 and
references therein).
If 1243+036 is also the case, the projected velocity field of the nebula
would show two components at different velocities each of which corresponds
to either front or rear surface of the expanding bubble or outflowing cone
(e.g., Heckman et al. 1990), rather than a single blueshifted component
as observed.
If the galaxy is surrounded by the dusty halo, which is often found
around HzPRGs (van Ojik et al. 1997), the redshifted emission
from the rear surface of the bubble may be attenuated due to the longer path
through the halo.
It seems also possible that the strong sky contamination at just red side of
the Ly$\alpha$ emission might make it very difficult to detect redshifted
counterparts of the blueshifted extended components.
Therefore the blueshifted and extended components at both sides of the nucleus
can be explained by the expanding bubble/outflowing cone model.
The top of the expanding bubble, if not blown out to form conical outflow
(e.g., Heckman et al. 1990), will show a single line-of-sight velocity.
If the superbubble around 1243+036 expands almost within
the sky plane, then
the observed velocity would be blueshifted from the systemic velocity
within the bubble, and it goes back closer to the systemic one at the tip.
Wider velocity width may be due to the kinematical disturbance at the shock
front of the bubble.
Since the superwind/superbubble model can reproduce the observed morphological and
kinematical properties of the nebula in qualitative ways, this model
seems more plausible than the model of irradiated disturbed matter by AGN.
Although we could not reject a possibility of the contribution of the AGN
excitation to the nebula, it seems very likely that the superwind activity
still plays an important role to bring the disturbed material around the galaxy
as observed.

\subsection{The superwind model}

We apply a simple superwind model to the nebula around 1243+036
to see whether the observed size and the kinematical conditions
can be reproduced quantitatively by the model.
To estimate order-of-magnitude values of the nebula size
and the expansion velocity, we apply a simple model of the evolution of
a single spherical superbubble.
The radius and the velocity of the expanding shell at time $t_8$
(in units of 10$^8$ yr) in the model are
\begin{equation}
r_{\rm shell}\sim 110 L^{1/5}_{\rm mech, 43} n^{-1/5}_{\rm H,
-5}t^{3/5}_{8}{\rm kpc}
\end{equation}
and
\begin{equation}
v_{\rm shell}\sim 650 L^{1/5}_{\rm mech, 43} n^{-1/5}_{\rm H,
-5}t^{-2/5}_{8}{\rm km~ s}^{-1}
\end{equation}
where $L_{\rm mech, 43}$ is the mechanical luminosity released
collectively from the supernovae in the central starburst in units of
$10^{43}$ erg s$^{-1}$, and $n_{\rm H, -5}$ is the average hydrogen
number density of the inter galactic medium (IGM) in units of
$10^{-5}$ cm$^{-3}$ (McCray \& Snow 1979; Koo \& McKee 1992a, 1992b;
Heckman et al. 1996).
Since the host galaxy is luminous in blue in the rest-frame of the galaxy
($M_{\rm B}=-25.4$; van Breugel et al. 1998), it seems to
be a massive elliptical galaxy with a mass $\sim 10^{12} M_{\rm \odot}$.
Assuming an instantaneous burst of star formation of $10^{11}M_{\rm \odot}$,
the mechanical luminosity is estimated to $L_{\rm mech}\sim 10^{44}$ erg
s$^{-1}$ over a time scale of $\gtrsim 10^{7}$ yrs (Leitherer et al. 1999)
\footnote
{Although the estimate of the mechanical luminosity depends strongly on
the choice of the shape of the stellar initial mass function (upper- and
lower-mass cut-offs and slope of the function) and metallicity as well as
the mass of the burst, and is difficult to be made accurately,
results of the model calculation are not sensitive to
the adopted mechanical luminosity since both $r_{\rm shell}$ and $v_{\rm shell}$
are proportional to just one fifth power of $L_{\rm mech, 43}$.
}.
The average hydrogen number density of the IGM is estimated following
Taniguchi \& Shioya (2000), and is $1.1 \times 10^{-4} h^{-2} = 4.4
\times 10^{-4}$ at $z=3.6$ in the adopted cosmology.
Substituting all these values into the equations, we find $r_{\rm
shell} \simeq 80 t^{3/5}_8$ kpc and $v_{\rm shell} \simeq 480 t^{-2/5}_8$
km s$^{-1}$.
Therefore both the observed size of the blueshifted extended nebula
($\simeq 30$ kpc in radius, corresponding to 4.5\arcsec)
and the maximum blueshift velocity ($-1400$
km s$^{-1}$) can be roughly reproduced if we assume $t_8 \simeq$ 0.1 -- 0.3,
which is comparable to the age of a nearby superwind nebula of Arp 220
(e.g., Heckman et al. 1996).
Hence the superwind model can be applicable to the blueshifted extended
nebula of 1243+036 in a quantitative way.

\section{Summary and Concluding Remarks}

We found an elongated giant nebula ($\sim 60$ kpc)
with blueshifted ($-1100$ -- $-1400$ km s$^{-1}$) and broad
($\simeq$ 1000 -- 1900 km s$^{-1}$ FWHM) Ly$\alpha$ emission
around a HzPRG 1243+036 at $z=3.6$.
The nebula can be recognized as a pair of expanding elongated bubbles or
outflowing bi-directional cones at both sides of the nucleus, being much
similar to some prototypical superwind galaxies in local universe in both
morphological and kinematical points of view.
We show that the superwind model can be applicable to the nebula to explain
the size and the kinematical properties of its bubble or cone-like structure.

In order to see such an evolved nebula associated with the superwind
activity at $z=3.6$, the onset of the starburst should be at
$z = 5 \sim 7$.
The is because it takes several $10^8$ yrs for the superwind to
blow out of the host galaxy since the onset of the burst (Arimoto \&
Yoshii 1987), and another a few $10^7$ yrs is required for the nebula
to expand to the observed size.
If this would be the case, it seems very likely that 1243+036 is in
the phase of the galactic wind after the initial burst for the galaxy
formation (Arimoto \& Yoshii 1987).

\acknowledgements
We thank L. Pentericci for her valuable comments and suggestions
to improve this paper.

% Figure 1
\figcaption{
Our $V$-band image (contours) and a HST F702W image of Pentericci et al.
(1999) (gray scale) are compared.
North is up and east is to the left.
Relative coordinates from the nucleus are shown in both RA and Dec.
The lower-left circle represents our effective seeing size after Gaussian
smoothing (1.1\arcsec~ FWHM).
The object at $\sim 2$\arcsec~ southwest of the nucleus may not be part of
the system (van Breugel et al. 1998).
Lower right contours seen in $B$-band image is of a nearby bright star.
}

%Figure 2
\figcaption{
A $V$-band image and Ly$\alpha$ spectrogram of 1243+036.
(a) $V$-band image after smoothing by the Gaussian convolution.
The effective seeing size after the smoothing size is 1.\arcsec1~
FWHM, and is shown by the lower-left circle.
Lower right emission is a part of the nearby bright star.
The image is rotated so that the slit (0.\arcsec8~ width, at PA=152$^\circ$)
is shown vertically on it (with two vertical lines at the center).
(b) One-dimensional $V$-band flux distribution along the slit.
(c) Ly$\alpha$ spectrogram after averaging over $4\times
4$ binned-pixel averaging.
Vertical axis is for spatial direction (upper side is toward northwest
and lower side is toward southeast), and horizontal axis is for
wavelength direction (right side is toward red).
The region contaminated by the strong sky emission at 5577\AA~ is not
shown, and is marked in blue.
(d) One-dimensional Ly$\alpha$ flux distribution along the slit.
}

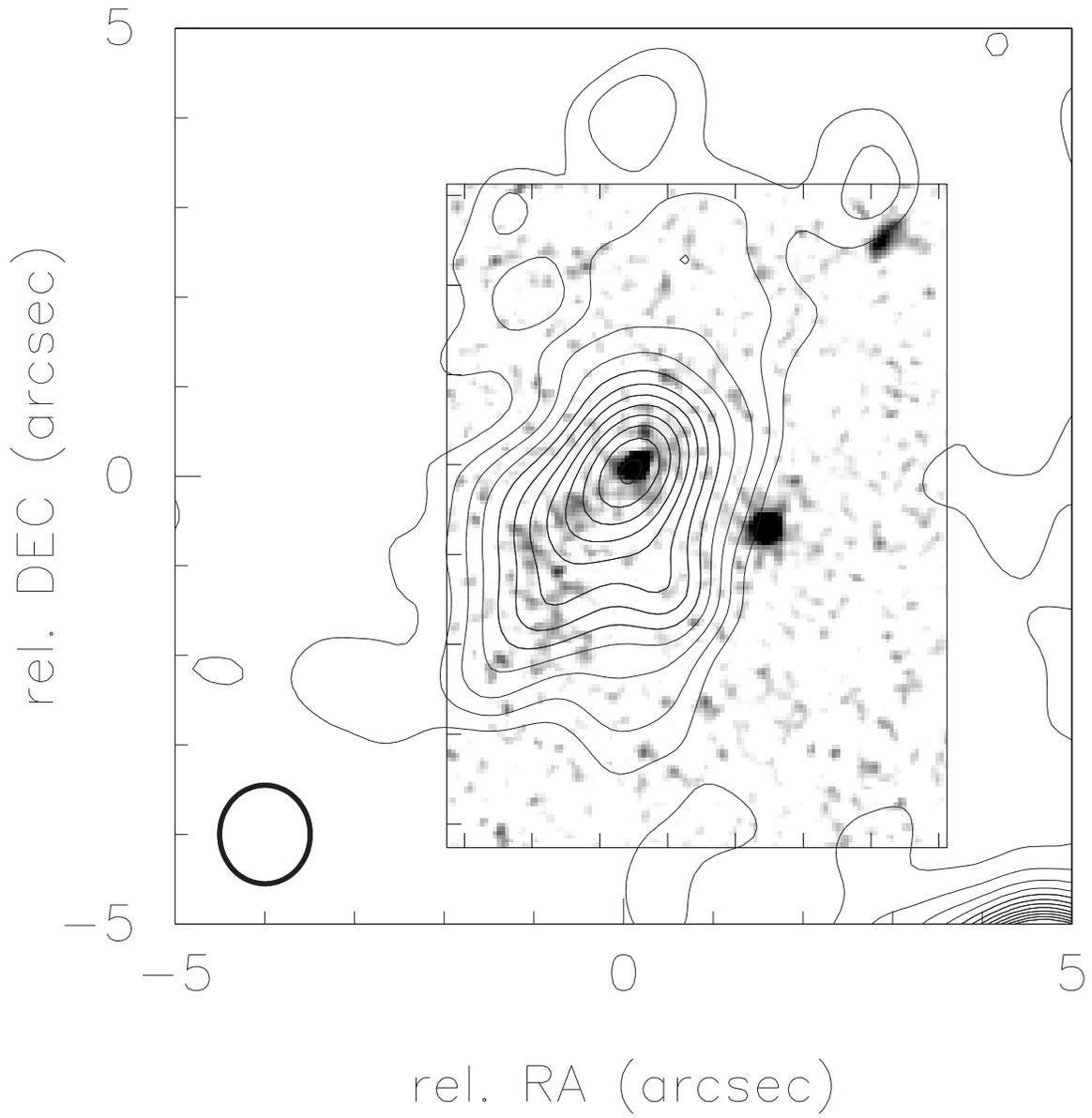
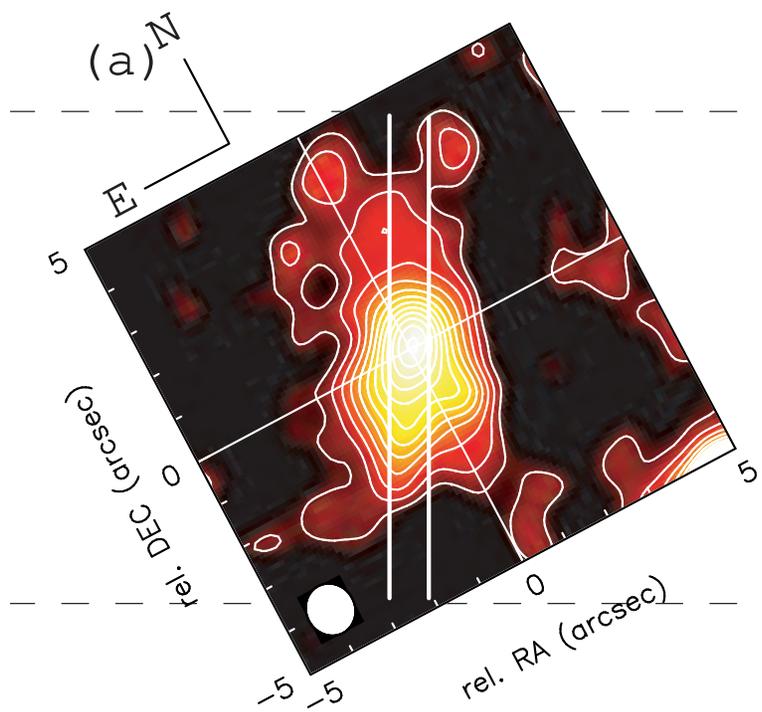
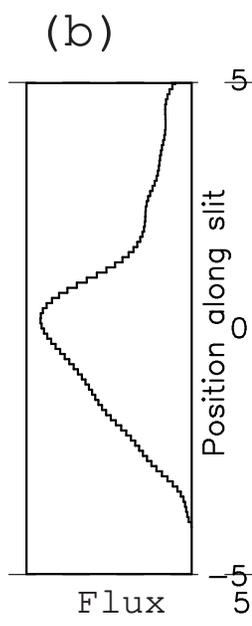
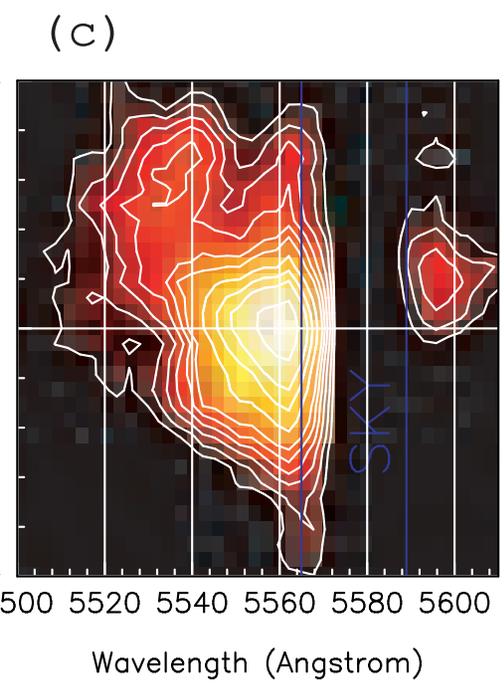
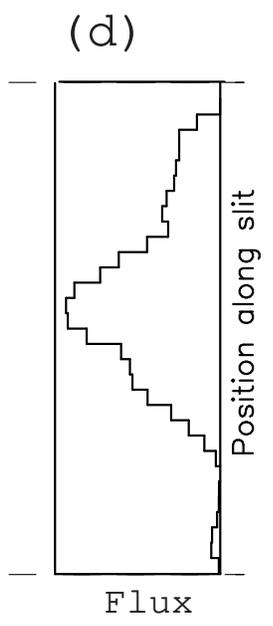
\end{document}